# Guidelines for the Search Strategy to Update Systematic Literature Reviews in Software Engineering

Claes Wohlin
*Department of Software Engineering*
*Blekinge Institute of Technology*
Karlskrona, Sweden
*claes.wohlin@bth.se*

Emilia Mendes
*Department of Computer Science*
*Blekinge Institute of Technology*
Karlskrona, Sweden
*emilia.mendes@bth.se*

Katia Romero Felizardo
*Department of Computing*
*Federal Technological University of Paraná*
Cornélio Procópio, Brazil
*katiascannavino@utfpr.edu.br*

Marcos Kalinowski
*Department of Informatics*
*Pontifical Catholic University of Rio de Janeiro*
Rio de Janeiro, Brazil
*kalinowski@inf.puc-rio.br*

**Abstract**

**Context**: Systematic Literature Reviews (SLRs) have been adopted within Software Engineering (SE) for more than a decade to provide meaningful summaries of evidence on several topics. Many of these SLRs are now potentially not fully up-to-date, and there are no standard proposals on how to update SLRs in SE.
**Objective**: The objective of this paper is to propose guidelines on how to best search for evidence when updating SLRs in SE, and to evaluate these guidelines using an SLR that was not employed during the formulation of the guidelines.
**Method**: To propose our guidelines, we compare and discuss outcomes from applying different search strategies to identify primary studies in a published SLR, an SLR update, and two replications in the area of effort estimation. These guidelines are then evaluated using an SLR in the area of software ecosystems, its update and a replication.
**Results**: The use of a single iteration forward snowballing with Google Scholar, and employing as a seed set the original SLR and its primary studies is the most cost-effective way to search for new evidence when updating SLRs. Furthermore, the importance of having more than one researcher involved in the selection of papers when applying the inclusion and exclusion criteria is highlighted through the results.
**Conclusions**: Our proposed guidelines formulated based upon an effort estimation SLR, its update and two replications, were supported when using an SLR in the area of software ecosystems, its update and a replication. Therefore, we put forward that our guidelines ought to be adopted for updating SLRs in SE.

*Keywords* — Systematic Literature Review Update, Systematic Literature Reviews, Software Engineering, Snowballing, Searching for evidence

## 1. Introduction

In 2004, Kitchenham et al. [20] argued for an Evidence-Based paradigm in Software Engineering (EBSE), to be mainly employed by "researchers interested in empirical software engineering and practitioners faced with decisions about the adoption of new software engineering technologies". EBSE's goals are to: "provide the means by which current best evidence from research can be integrated with practical experience and human values in the decision-making process regarding the development and maintenance of software", and also to encourage the use of Systematic Literature Reviews (SLRs) to obtain such current best evidence.

Such call to arms prompted the Software Engineering (SE) community to publish SLRs, leading to more than 430 SLRs published within the period from January 2004 to May 2016 [19], [24], [54] and [4]. Despite such a large number of published SLRs, Mendes et al. [42] identified only 20 updated SLRs, within the period 2006 to 2018. The findings show that many SLRs in SE are potentially not fully up to date, thus influencing our current aggregated understanding of the state-of-the-art. If we also take into account that primary studies may be getting old, e.g., studies that used a particular technology that is not suitable any longer, the situation is worsened.

There are a few studies on the topic of SLR updates in SE (see Section 2). However, there has been no study that has systematically compared different search approaches to make recommendations on how to identify new evidence when updating SLRs in SE. It is essential to note that the starting point for an SLR update are different from conducting the original SLR, i.e. we already have an SLR. Thus, we need to decide whether to perform a replication of the search strategy in the original SLR, although with a different time window, or to use the information available in the original SLR to conduct the search for new papers for the update. Coming up with guidelines for an SLR update search strategy was the goal and the main contribution of our research presented in [43], and this paper is an extension of [43]. The formulation of the guidelines is also presented herein, together with an extensive evaluation of the guidelines.

Note that the issue of when to update an SLR in SE has also been addressed and is detailed elsewhere [42]. In this context, it is essential first to identify whether or not there is a need for an update, i.e. when to update, before updating an SLR. However, once the need for an update is identified, it is crucial to maximizing the potential of the update by using the best possible search strategy.

In [43], we compared and discussed outcomes from applying different search strategies (e.g., database search and forward snowballing processes) using an SLR on the topic of effort estimation. We employed the results from both an SLR update and its replications, where such replications used different search strategies to identify primary studies. We also looked for an SLR update that had replications carried out by, if possible, different authors. Such diversity of authors was essential to reduce as much as possible any bias when applying different methods to identify primary studies. Out of the twenty SLR updates found [42], only one met our criteria. At the expense of a better conclusion validity, our focus was not to carry out a formal experiment comparing different ways to identify primary studies. We wanted to base our suggestions upon already existing evidence from replications of SLR updates in SE.

Furthermore, both the updated SLR and its replications were conducted and reported at different times, which is not optimal. However, if we were to replicate the searches and examine the studies at the same time, we would potentially bring enough bias into the process to make findings less trustworthy. Based on the SLR, its update and the replications, we proposed guidelines for the search strategy for updating SLRs. The guidelines suggest the use of a single iteration forward snowballing with Google Scholar, employing as seed set the original SLR and its primary studies.

However, the guidelines needed to be further evaluated. One possible way forward would be replicating a published SLR update. Such replication is one of the main contributions of this paper. Once selecting a suitable SLR for the replication, an SLR update was conducted using the guidelines and the outcome is compared with the results in the published SLR update. Including this additional replication, our guidelines were extensively evaluated and based on the investigation, and it was concluded that following the guidelines concerning the search strategy for updating SLRs is a better alternative than doing a database search.

The remainder of this paper is organised as follows. In Section 2, we introduce related work. In Section 3, the research design is presented. The SLR selected to formulate the proposed guidelines for updating SLRs and the SLR chosen to be used in the evaluation of the guidelines are detailed in Section 4. Section 5 puts forward the formulation of the guidelines based on an analysis of different search strategies for updating SLRs. The guidelines are evaluated by conducting and comparing a new SLR update with a published SLR update in Section 6. In Section 7, the threats to validity are discussed for both the formulation and evaluation of the guidelines. Finally, Section 8 concludes the work concerning how to update SLRs, and in particular related to the search strategy.

## 2. Related work

Challenges concerning updating SLRs have been identified, and some different aspects have been researched. It includes, for example, the research by Dieste et al. [7] and Ferrari et al. [10] focusing on the development of processes to support updating SLRs in SE. However, their processes do not focus on "how" to best search for new evidence, i.e. the search strategy for updating SLRs.

Felizardo et al. [9] proposed an approach based on Visual Text Mining (VTM), which supports the selection of new evidence to update an SLR using the studies included in the original SLR as a starting point. The research presented here is different since it does not focus on the use of an analytical technique that employs visualization (e.g. VTM). Our research, including the guidelines and their evaluation, is based on a detailed comparison of different search strategies. The recommendations are based on an SLR update and its replications; and whether different approaches lead to significant differences in the set of included studies or the SLR conclusions. Furthermore, we provide an additional evaluation of the recommendations.

Silva et al. [53] evaluated different search sources (specific – the IEEE Xplore database and generic – the Google Scholar indexing service) for supporting secondary studies' updates. IEEE Xplore is judged not sufficient to identify most studies; conversely, Google Scholar seems adequate. Despite their focus on the selection of databases/services, they neither compare different search approaches, nor whether different approaches lead to significant differences in the set of included studies or the SLR conclusions.

Rodriguez et al. [51] reported lessons learned considering their experience in updating SLRs. Some of these lessons-learned are: (i) to adopt software tools to support the updating process; (ii) to provide as much information as possible about the SLR being updated; (iii) to involve some of the authors from the SLR being updated; and (iv) to reuse the protocol from the SLR being updated. Nepomuceno and Soares [47] extended Rodriguez et al.'s work [51], by asking researchers about the lessons documented by Rodriguez et al. [51] and reached similar conclusions.

Similar to Rodriguez et al. [51], and Nepomuceno and Soares work [47], in this paper, we also provide lessons learned when updating SLRs. However, our primary focus is to use evidence from a detailed comparison between different search strategies to provide concrete recommendations on how to search for new evidence when updating SLRs in SE. Moreover, we evaluate the guidelines on a separate SLR and its update. To date, there has been no such detailed study in SE concerning the search strategy for updating SLRs.

## 3. Research design

The research team for formulating the guidelines concerning the search strategy for updating systematic literature reviews emerged from a joint interest in the topic. The four authors of this paper have all been involved in conducting SLRs. Two of the authors have also been involved in updating SLRs. Finally, all four authors have been involved in replicating updated SLRs to compare different search strategies.

### 3.1 Formulating the guidelines

Based on twenty possible SLR candidates having both an SLR and an update, an SLR was selected to formulate the guidelines. The selection criteria concerned the suitability of the SLR update for our purposes and the collective expertise of the authors. The main criterion concerned having both an update and replications using different search strategies. The selected SLR was an SLR looking at papers comparing cross-company (CC) vs within-company (WC) effort estimation. The original SLR was published in 2006–2007 [22] and [23], and an update was published in 2014 [37]. The update was replicated twice [60] and [8] using different search strategies. Thus, this set of updates provided an excellent opportunity to compare and hence recommend a search strategy for updating SLRs. The three papers used in the formulation of the search strategy guidelines are the updated SLR [37] (henceforth denoted SLR-update) and the two replications by respectively Wohlin [60] (henceforth denoted SLR-update-R1, where R refers to replication) and Felizardo et al. [8] (henceforth denoted SLR-update-R2). These three papers are all the updates of the original SLR, although applying different search strategies.

After selecting the SLR, five research questions (RQs) concerning how best to search for evidence when updating an SLR were formulated, as follows:

- **RQ1:** Were all studies in the superset retrieved by the different searches employed by the SLR-update, SLR-update-R1, and SLR-update-R2?
- **RQ2:** If not all studies in the superset were included, which of the studies were selected/included by the SLR-update, SLR-update-R1, and SLR-update-R2, respectively?
- **RQ3:** Concerning forward snowballing,
    - **RQ3.1:** Do different seed sets lead to differences in the set of included studies and the SLR conclusions?
    - **RQ3.2:** Do different processes lead to differences in the set of included studies and the SLR conclusions?
- **RQ4:** Were there differences between the conclusions from the SLR-update, SLR-update-R1, and SLR-update-R2, when compared to the conclusions that would have been obtained using the superset?
- **RQ5:** What are the differences, if any, between the conclusions using the superset, and those in the original SLR?

RQ1 focuses upon the search strategy employed in the SLR update and its two replications. We consider the search strategy separately from the selection of primary studies since we want to investigate the cause(s) for not all studies being included in all three updates, i.e. the SLR-update and its two replications. We refer to all papers found by the three updates as the superset. Thus, the superset is the union of the papers in the SLR-update and the two replications. For RQ2, we are interested in which papers in the superset each of the three updates identifies. Although RQ2 has been addressed in the two replications of the update, we cover it here since it is highly relevant to understand how to search for new evidence when updating SLRs. RQ3 relates to differences in conducting forward snowballing, i.e. identifying citations to a paper. The term seed set is used to denote the set of papers used to start the forward snowballing. RQ4 and RQ5 concern the conclusions based on the included papers.

Further details regarding the selected SLR, its update and the two replications are presented in Section 4.1. The results relating to the research questions are presented in Section 5.1. Based on the results, guidelines for updating SLRs are proposed in Section 5.2.

## 3.2 Evaluating the guidelines

The outcome in the form of the proposed guidelines needed to be further evaluated. Thus, an evaluation study was designed. The evaluation was conducted on a separate SLR.

To select the SLR to be used in the evaluation, we applied a systematic approach. The basis for the strategy was the requirement that the SLR needed to have both an original SLR and an updated SLR, which should have been updated using a database search. Furthermore, we also applied additional criteria to select the candidate SLRs to use in the evaluation.

The following criteria were identified and used in the selection procedure:

- The original update did not use forward snowballing.
- The primary studies in the SLR need to be easily identifiable since they are going to be used for the forward snowballing.
- The selected SLR should, if possible, be in areas where the researchers conducting the update had sufficient knowledge to make an update.

- Given that the SLR used to formulate the search strategy for updating SLRs included the authors of this paper, it is preferred that in this evaluation step the authors were not co-authors of neither the original SLR nor the updated SLR of the selected SLR.

The candidate SLRs, and their updates were identified from [42]. The details on how the SLR was selected are provided in Section 4.2.

Once selecting a suitable SLR, an SLR update was done using the guidelines. It included analysing all citations to the original SLR and its primary studies and deciding on inclusion or exclusion of the papers citing them. The analysis resulted in identifying a set of papers, to be compared with the papers identified in the published updated SLR.

The objective was to perform a new SLR update using the proposed guidelines and then compare the results with those provided in the already published update of the SLR. Based on the objective, the research questions are as follows:

- **RQ6:** How do the proposed guidelines perform in terms of identifying papers to be included in the updated SLR, in comparison to the published SLR update?
- **RQ7:** To what extent the papers found in our SLR update have also been found when conducting the first SLR update?
- **RQ8:** Which of the two SLR updates identifies most papers focusing on research on the topic of the selected SLR?

To address the research questions, we defined an evaluation process. The process for conducting case studies inspired the process used here, i.e. the five main steps are taken from the research process for conducting case studies as suggested by Runeson et al. [52]. For some of the steps, substeps are identified. The evaluation process is as follows:

1. **Design:** The design involved selecting an SLR and applying the proposed guidelines for updating this SLR. To enable a comparison, we needed to select a suitable SLR to use in the evaluation. The selection of an SLR was based on a set of criteria.
2. **Preparation:** The conduct of the SLR update was distributed. The authors were based in three different locations in different time zones, and hence the communication was handled primarily over email. This decision led to high requirements on the preparation and the coordination of the research study. The first author prepared Excel sheets to support the evaluation. Furthermore, to ensure alignment, three important activities were conducted. All authors did a search in relation to the selected SLR, and the outcomes were compared to ensure synchronization concerning the retrieval of papers. The alignment continued with all authors individually looking at the inclusion of studies. The findings were compared, and different opinions were discussed and resolved. Finally, based on the alignment, it was decided to work in teams of two persons to make the work more efficient.
3. **Collecting data:** A forward snowballing was done on the original SLR and on all of its primary studies. Inclusion of papers in our updated SLR was based on the criteria documented in the published SLR update.
4. **Analysis:** An analysis was conducted to address the research questions. The comparison was made concerning the included papers and the coverage of the topic of the SLR.
5. **Reporting:** Based on the analysis, the observations from the evaluation were reported in relation to the guidelines for updating SLRs. Reflections on the proposed guidelines are provided.

The different steps in the evaluation process are detailed later in the paper. Further details concerning the SLR selection are presented in Section 4.2, while steps 2–5 are described in one subsection each in Section 6.

# 4. Selection of SLRs for formulating and evaluating the guidelines

## 4.1 SLR for formulating the guidelines

The SLR update and its two replications, used to formulate the guidelines, relate to an SLR on the topic of cross-company (CC) vs within-company (WC) effort estimation. The original SLR is, and its results were published as a conference paper [22] and later as a journal paper [23]. The list of primary studies included in the original SLR is provided in Table 1 and given a study identity (SID).

The primary studies are sorted based on the findings reported. The wording "Significantly different" in Table 1 and Table 2 means that the authors of the primary study carried out a statistical significance test to compare the predictions obtained via models built using within- and cross-company data, and these predictions were statistically significantly different. The inconclusive studies refer to studies that for various reasons, did not conduct a statistical significance test of results, for example, due to some missing data or information.

Table 1. STUDIES INCLUDED IN THE ORIGINAL SLR.

| SID | Authors | Ref. | Year |
|---|---|---|---|
| **CC model NOT significantly different from WC model** | | | |
| S2 | L.C. Briand et al. | [2] | 1999 |
| S3 | L.C. Briand et al. | [3] | 2000 |
| S6 | I. Wieczorek and M. Ruhe | [58] | 2002 |
| S10 | E. Mendes et al. | [41] | 2005 |
| **CC model significantly different from WC model** | | | |
| S4 | R. Jeffery et al. | [16] | 2000 |
| S5 | R. Jeffery et al. | [15] | 2001 |
| S8 | B. A. Kitchenham and E. Mendes | [21] | 2004 |
| S9 | E. Mendes and B. A. Kitchenham | [38] | 2004 |
| **Inconclusive** | | | |
| S1 | K. Maxwell et al. | [34] | 1999 |
| S7 | M. Lefley and M. Shepperd | [28] | 2003 |

This original SLR was updated once [37] (SLR-update), and later replicated twice by studies investigating different search strategies ([60] and [8]) and whether they would retrieve the same primary studies as the SLR-update. These two replications are named henceforth as SLR-update-R1 and SLR-update-R2, respectively. Together the three updates identified a superset containing 15 studies published until the end of 2013, shown in Table 2. Please note that papers and studies are used interchangeably, although adapted to the context since papers identified in an SLR are often referred to as primary studies.

Note that there were several studies where the results contrasted, depending on the prediction models compared. It explains why we have some studies (S14, S15, S16, and S20) shown under different categories in Table 2. The SLR-update should have identified, except for S24, all 14 studies (explanation given later), and both replications should have identified all 15 studies. However, this was not the case, as shown in Figure 1. The results did not entirely agree. Such findings motivated us to investigate further the issue of searching for evidence when updating SLRs; this is reflected in our research questions. Some further details on the SLR-update and its two replications are provided next.

The SLR update [37] was published in 2014 and had the participation of two of the four co-authors of this paper – Mendes and Kalinowski. It used the same search string and protocol as in the original SLR. In total, the SLR-update identified 11 primary studies, and missed studies S22 to S25. The first replication of SLR-update-R1 [60] was published in Q2 2016 and was authored by one of the four co-authors of this paper – Wohlin. The seed set Wohlin used to replicate the SLR-update contained 12 sources: the two papers that detailed the original SLR [22] and [23] plus the ten primary studies included in the original SLR (listed in Table 1). Wohlin analysed the citations provided by Google Scholar to each of the 12 sources as a means to identify possible additional studies without any iteration. He followed his proposed method, as described in [59]. A total of 12 studies were selected, where a selected study means that the author(s) decide to include the study.

Table 2. LIST OF 15 STUDIES IN THE SUPERSET

| SID | Authors | Ref. | Year |
|---|---|---|---|
| **CC model NOT significantly different from WC model** | | | |
| S13 | C. Lokan and E. Mendes | [30] | 2008 |
| S14 | E. Mendes and C. Lokan | [40] | 2008 |
| S15 | E. Mendes et al. | [35] | 2008 |
| S16 | C. Lokan and E. Mendes | [29] | 2009 |
| S18 | E. Kocaguneli and T. Menzies | [27] | 2011 |
| S20 | F. Ferrucci et al. | [12] | 2012 |
| S22 | R. Premraj and T. Zimmermann | [48] | 2007 |
| S23 | E. Kocaguneli et al. | [26] | 2010 |
| **CC model significantly different from WC model** | | | |
| S11 | C. Lokan and E. Mendes | [31] | 2006 |
| S12 | E. Mendes et al. | [36] | 2007 |
| S14 | E. Mendes and C. Lokan | [40] | 2008 |
| S15 | E. Mendes et al. | [35] | 2008 |
| S16 | C. Lokan and E. Mendes | [29] | 2009 |
| S17 | E. Mendes and C. Lokan | [39] | 2009 |
| S20 | F. Ferrucci et al. | [12] | 2012 |
| S21 | L. L. Minku and X. Yao | [44] | 2012 |
| **Inconclusive** | | | |
| S19 | O. Top et al. | [55] | 2011 |
| S24 | E. Kocaguneli et al. | [25] | 2013 |
| S25 | F. Ferrucci et al. | [11] | 2009 |

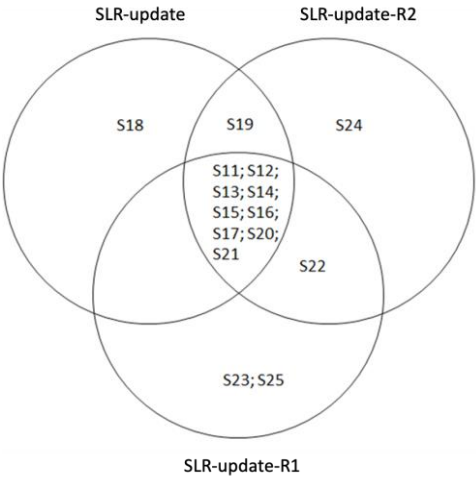

Figure 1. The 15 studies in the superset organised by studies (SLR update and its replications).

In comparison to the SLR update, it did not include S18, S19 and S24. However, it included three additional studies (S22, S23 and S25) not included by the SLR-update. Finally, the second replication of the SLR-update U1 – SLR-update-R2 [8], was published in Q3 2016 and was co-authored by three of the four co-authors of this paper – Felizardo, Mendes and Kalinowski. They performed forward snowballing on a seed set containing the ten primary studies included in the original SLR. Citations were identified via search engines, such as IEEEXplore and ACM, instead of using Google Scholar. Four iterations were carried out until reaching a saturation point. A total of 172 studies were found, of which 12 were selected for inclusion. The approach identified all the studies included in the SLR-update, except for one – S18; and also identified two studies (S22 and S24) not included in SLR-update. In comparison with SLR-update-R1, it did not include two studies (S23 and S25) and included two studies not included by SLR-update-R1 (S19 and S24). It should be noted that SLR-update-R2 originally included 13 papers. However, during the writing of this paper, we noticed that one of the papers (called N3 in SLR-update-R2) was incorrectly included. The reason is a miscommunication between two of its authors.

In summary, the two replications of the SLR-update jointly found another four studies that were not included by the SLR-update (S22, S23, S24 and S25). Furthermore, Table 3 provides an overview of SLR-update-R1 and SLR-update-R2, showing the number of iterations, studies found, studies included, unique studies identified by each review and studies that are common with the SLR-update.

Table 3. SUMMARY OF RESULTS FOR THE TWO REPLICATIONS.

| Characteristics | SLR-update-R1 [60] | SLR-update-R2 [8] |
|---|---|---|
| Seed set | 12 (10 + 2 SLRs) | 10 |
| Iterations | 1 | 4 |
| Citations analysis of studies found | 1018 | 172 |
| Studies included | 12 | 12 |
| In common with the SLR-update | 9 | 9 |
| Unique studies | 2 | 1 |

## 4.2 SLR for evaluating the guidelines

The first step of our evaluation was to select a suitable SLR, requiring both an original SLR and an update of the original SLR. We wanted an SLR that had a well-motivated update, and we used the results from a separate line of research that investigated how to determine when the update of SLRs is well-motivated [42]. In [42], we identified six SLRs that are well-motivated to update and that have been updated (see Table 4).

Table 4. CANDIDATE SLRS FOR THE EVALUATION

| | *Updated SLR* | | | | *Original SLR* | |
|---|---|---|---|---|---|---|
| **SID** | *Research Topic* | *Year* | *Search span* | *Timespan between SLRs* | *Search span* | *Year* |
| **SLR1** | Computer Games and Serious Games [5] and [1] | 2016 | 2009 – 2014 | 4 years | 2004 – 2009 | 2012 |
| **SLR2** | Effort Estimation [22], [23] and [37] | 2014 | 2006 – 2013 | 7 and 8 years | 1990 – 2006 | 2006 |
| | | | | | 1990 – 2006 | 2007 |
| **SLR3** | Effort Estimation [56] and [6] | 2018 | 2014 – 2017 | 4 years | 2001 – 2013 | 2014 |
| **SLR4** | Maintainability Prediction [49] and [50] | 2012 | 2008 – 2010 | 3 years | 1985 – 2008 | 2009 |
| **SLR5** | Software Ecosystem [33] and [32] | 2016 | 2007 – 2014 | 3 years | 2007 – 2012 | 2013 |
| **SLR6** | Agile practices [13], [14] and [57] | 2018 | 2010 – 2016 | 6 and 8 years | 1999 – 2009 | 2010 |
| | | | | | 1999 – 2009 | 2012 |

As can be seen from Table 4, two SLRs (SLR2 and SLR6) include both an original conference publication and an extended journal version of the original SLR. In both cases, the extended journal version, which is published later, covers the same search span used in the conference paper version. SLR2 is the SLR used in Section 4.1, and hence it cannot be selected at this stage. To decide upon which of the other SLR candidates to choose, we formulated a set of criteria to make an informed decision, as described in Section 3.2.

The SLR referred to as SLR4 used forward snowballing to update the original SLR, and hence SLR4 is not suitable for our intended comparison. It leaves four candidate SLRs. The four candidate SLRs all provide sufficient information to find and use the primary studies when doing forward snowballing, and hence the number of candidates is still four. The authors' knowledge concerning computer games and serious games is limited, and therefore SLR1 is not a suitable candidate for that reason. It leaves three candidate SLRs. Emilia Mendes is a co-author of the original SLR for SLR4, and Claes Wohlin is a co-author of the original SLR for SLR6. It leaves only SLR5 as a candidate, and hence SLR5 was selected for replicating the update using forward snowballing and accordingly evaluate the proposed guidelines.

SLR5 is in the area of software ecosystems. The original SLR was published in 2013 [33], and covering publications in the time interval 2007–2012. The update was published three years later and covered publications from 2007–2014 [32], and hence it is not an actual update since it revisits the period covered by the original SLR. Thus, the papers included in the update needed to be filtered to the search span 2012–2014. Moreover, papers from 2012 in the original SLR also need to be removed.

The selected original SLR includes 90 primary studies. The inclusion criteria should be the same as in the updated SLR [32]. The inclusion criteria are straightforward for this SLR:

- Papers should include either of the following wordings: "software ecosystem" or "software ecosystems".
- Papers should be written in English.

Moreover, according to the updated SLR, the following publications were excluded: books, short articles (less than two pages), conference keynotes, and extended abstracts. When it comes to books, we interpreted the formulations such that book chapters should be excluded. In the original SLR [33], some additional details are provided concerning the location of the wordings above. It is assumed that they have also been applied in the updated SLR. The assumption is based on the fact that the main author is the same for both the original SLR and the updated SLR. In the original SLR, it is stated that the abovementioned keywords should be in the title, abstract, or keywords. The original SLR also stresses that papers should be peer-reviewed, which we assume is covered by the exclusion of specific type of publications.

Furthermore, the original SLR describes how papers with composite expressions, including the words "software" and "ecosystem" are included, for example, software-intensive ecosystems. Once again, we assume that this was also applied in the updated SLR. Finally, the original SLR includes a formulation stating that the selected papers could include software ecosystems either as a main or a secondary research area. This formulation cannot be found in the update. When it comes to this criterion, we have chosen to go with the inclusion criteria documented in the updated SLR, i.e. we do not make any judgment in the selection step concerning software ecosystems as a research area. However, we do look at the papers included when comparing the updated SLR with our replication using the recommendations abovementioned.

It is essential to describe the search strategy applied in the updated SLR to understand the comparison between the updated SLRs. According to our interpretation, and using the original SLR as further input, the updated SLR's search strategy is as follows:

- Search in five databases using the keywords "software ecosystem" or "software ecosystems". The following five digital libraries were searched: ACM Digital Library, IEEE Explore, SpringerLink, ScienceDirect and Web of Science.
- Include papers from the International Workshop on Software Ecosystems and the International Workshop on Software Ecosystem Architectures that meet the inclusion criteria.
- Include articles from two special issues published in the Journal of Systems and Software 85(7) and the Journal of Information and Software Technology 56(11) that meet the inclusion criteria.

# 5. Formulating the guidelines

## 5.1 Results

Five research questions were formulated, as described in Section 3.1. The responses to the research questions formed the basis for formulating the guidelines for the search strategy concerning updating SLRs. The outcome in relation to the research questions is as follows:

***RQ1:*** *Were all studies in the superset retrieved by the different searches employed by the SLR-update, SLR-update-R1, and SLR-update-R2?*

Table 5 shows the papers retrieved by each of the three studies. A retrieved study is found by the search strategy, while a selected study implies that the author decided to include the retrieved study in the analysis. Thus, we may have differences between retrieved relevant papers and selected papers, even when a paper at the end is judged as belonging to the superset. The only search strategy that retrieved all 15 studies was the search carried out in SLR-update-R1, using as seed set the two references to the original SLR, and all the ten primary studies included in the original SLR.

When it comes to the SLR-update, paper S25 was not indexed by any of the search engines, so it would not be possible to find it either via a normal database search or using the ACM and IEEE citation search mechanisms. Paper S24 was presented at a conference in October 2013, which probably explains why it was not found during the database searches carried out in the SLR-update early November 2013.

Table 5. PAPERS RETRIEVED BY THE DIFFERENT SEARCH STRATEGIES APPLIED IN THE UPDATES COMPARED TO THE SUPERSET.

| Superset | Retrieved by | | |
|---|---|---|---|
| | SLR-update | SLR-update-R1 | SLR-update-R2 |
| **S11** | √ | √ | √ |
| **S12** | √ | √ | √ |
| **S13** | √ | √ | √ |
| **S14** | √ | √ | √ |
| **S15** | √ | √ | √ |
| **S16** | √ | √ | √ |
| **S17** | √ | √ | √ |
| **S18** | √ | √ | × |
| **S19** | √ | √ | √ |
| **S20** | √ | √ | √ |
| **S21** | √ | √ | √ |
| **S22** | √ | √ | √ |
| **S23** | √ | √ | × |
| **S24** | × | √ | √ |
| **S25** | × | √ | × |

**Type of search**: SLR-update (Search string); SLR-update-R1 (Forward Snowballing – Google Scholar); SLR-update-R2 (Forward Snowballing – IEEEXplore and ACM).
**Legend**: √ – Yes; × – No

Finally, in comparison with SLR-update-R2, papers S18 and S23 were not retrieved by the ACM or the IEEE citation searches, despite being indexed by both search engines.

***RQ2:*** *If not all studies in the superset were included, which of the studies were selected/included by the SLR-update, SLR-update-R1, and SLR-update-R2, respectively?*

Except for SLR-update-R2, some papers were retrieved, although not included in both the SLR-update and SLR-update-R1 (Table 6). Both the SLR-update and SLR-update-R1 had to look over a large number of titles and abstracts. We hypothesize that it is a likely reason for the false-negative results. Moreover, SLR-update-R1 was conducted by a sole researcher, which may also affect the results. Note that we also checked (using Google Scholar) how many papers would have been retrieved using only the original SLR as a seed set; it would have retrieved all papers, except for S19. It shows that using the original SLR solely as a seed set would not have been sufficient to retrieve all the 15 studies in the superset.

Please note that Figure 1 shows the primary studies included, whereas Tables 5 and 6 detail respectively papers retrieved by the searches and whether they were included or not in the SLR update or its replications.

Table 6. STUDIES INCLUDED IN THE SLR-UPDATE, SLR-UPDATE-R1, AND SLR-UPDATE-R2.

| Study | | SLR-update | SLR-update-R1 | SLR-update-R2 |
|---|---|---|---|---|
| **Study included** | S11 | √ | √ | √ |
| | S12 | √ | √ | √ |
| | S13 | √ | √ | √ |
| | S14 | √ | √ | √ |
| | S15 | √ | √ | √ |
| | S16 | √ | √ | √ |
| | S17 | √ | √ | √ |
| | S18 | √ | ? | × |
| | S19 | √ | ? | √ |
| | S20 | √ | √ | √ |
| | S21 | √ | √ | √ |
| | S22 | ? | √ | √ |
| | S23 | ? | √ | × |
| | S24 | × | ? | √ |
| | S25 | × | √ | × |

Legend: √ – Included; ? - retrieved but not included; × – Not retrieved

*RQ3: Concerning forward snowballing.* There were two specific sub-questions. The answers to these questions follow.

*RQ3.1: Do different seed sets lead to differences in the set of included studies and the SLR conclusions?*

Here we compare three different choices of seed sets (SLRs only; SLRs + primary studies; and primary studies only). Hence, we added one more possible seed set, i.e. using solely the two papers that describe the original SLR. This new seed set is searched using IEEE Xplore + ACM (to complement the seed set used in SLR-update-R2 and to be equivalent to the seed set used in SLR-update-R1), and also searched using Google Scholar (to assess whether SLR-update-R1 would have the same papers using only the original SLR as a seed set). The different seed sets are denoted SS1-SS4, as shown in Table 7. The table shows the total number of papers (titles and abstracts) that had to be checked and the list of included papers. If SLR-update-R2 had also included in its seed set the two papers describing the original SLR, it would have identified 13 new studies, rather than 12 (S23 would be the 13th paper).

Concerning the differences in the SLR conclusions when having different seed sets, we have the following:

- S14: The cross-company model is NOT significantly different from the within-company model (not retrieved by SS1)
- S18: The cross-company model is NOT significantly different from the within-company model (not retrieved by SS1, SS2 and SS4, and not included by SS3)
- S23: The cross-company model is NOT significantly different from the within-company model (not retrieved by SS4)
- S22: The cross-company model is NOT significantly different from the within-company model (not retrieved by SS1)
- S19: Inconclusive results (not retrieved by SS1 and SS2, and not included by SS3)

- S24: Inconclusive results (not retrieved by SS2 and not included by SS3)
- S25: Inconclusive results (not retrieved by SS1 and SS4)

Table 7. SEED SET AND THEIR RESPECTIVE SET OF INCLUDED STUDIES.

| SS# | SS1 | SS2 | SS3 | SS4 |
|---|---|---|---|---|
| **Seed set** | Original SLR only – 2 papers: [22] and [23] IEEE Xplore + ACM) | Original SLR only – 2 papers: [22] and [23] Google Scholar | Original SLR + Primary studies Google Scholar | Primary studies only IEEE Xplore + ACM |
| **# Studies Retrieved** | 114 | 224 | 1018 | 172 |
| **Studies Included (Superset = 15 studies)** | (10) S11; S12; S13; S15; S16; S17; S20; S21; S23; S24 | (12) S11; S12; S13; S14; S15; S16; S17; S20; S21; S22; S23; S25 | (12) S11; S12; S13; S14; S15; S16; S17; S20; S21; S22; S23; S25 | (12) S11; S12; S13; S14; S15; S16; S17; S19; S20; S21; S22; S24 |

There were four conclusive studies (S14, S18, S22 and S23) either not retrieved or not included by at least one of the seed sets, all providing evidence supporting cross-company models NOT being significantly different from within-company models. However, all these seed sets included at least four studies (S13, S15, S16 and S20) that also provided evidence supporting cross-company models NOT being significantly different from within-company models. Therefore, the overall SLR conclusions, based on each of the four seed sets, are well aligned and almost similar.

*RQ3.2: Do different processes lead to differences in the set of included studies and the SLR conclusions?*

Here we compare two different processes – Alternative 1: using Google Scholar and no iteration, which assumes that SLRs or primary studies are cited; and Alternative 2: using IEEEXplore and ACM, and iteration until saturation is reached.

To answer this question, we merged the results from the first and fourth seed sets (see Table 7) into Alternative 2 (Table 8).

In doing so, we acknowledge that any update to an existing SLR that uses forward snowballing should include in its seed set not only the primary studies included in that SLR but also the reference(s) to the published SLR. Furthermore, to provide a good discussion, it is also essential to consider the ideal results that could have been achieved using each of the two processes, i.e. if all the papers that should have been selected were selected by the researchers carrying out those tasks. In other words, if there had been no human error during the filtering process in Alternative 1 (or different judgment given that SLR-update-R1 was conducted by a single researcher), all the 15 papers would have been retrieved (see Table 8); this does not apply to Alternative 2, as it did not retrieve S18 and S25 in the first place.

Concerning the differences in the SLR conclusions (considering only the studies selected by the researchers, as per Table 8), we have the following:

- S18: The cross-company model is NOT significantly different from the within-company model (not included by Alternative 1 and not retrieved by Alternative 2)
- S19 and S24: Inconclusive results (not included by Alternative 1)
- S25: Inconclusive results (not retrieved by Alternative 2)

There was only one conclusive study (S18) not included by Alternative 1 and not retrieved by Alternative 2, suggesting that, at least within this context, there would not have been any significant changes to the SLR results with using either Alternative 1 or Alternative 2.

Table 8. SEED SET AND ITS RESPECTIVE SET OF INCLUDED STUDIES

| Seed set | Studies Included (Superset = 15 studies) |
|---|---|
| Alternative 1 – Original SLR + Primary studies (Google Scholar) + no iteration | (12) – S11; S12; S13; S14; S15; S16; S17; S20; S21; S22; S23; S25 (retrieved but did not include S18, S19 and S24) |
| Alternative 2 – Original SLR + Primary studies (IEEEXplore + ACM) + saturation | (13) – S11; S12; S13; S14; S15; S16; S17; S19; S20; S21; S22; S23; S24 (did not retrieve S18 and S25) |

*RQ4: Were there differences between the conclusions from the SLR-update, SLR-update-R1, and SLR-update-R2, when compared to the conclusions that would have been obtained using the superset?*

When we discard the inconclusive results, we can see that there are only three studies that were not included by the SLR update and the replications, namely S18, S22 and S23. S18 and S23 show results that are also recurrent in many other studies (WC superior or CC and WC not significantly different). However, S22 presents one of the only two results to date where CC showed superior accuracy in effort estimation to WC. SLR-update-R1 and SLR-update-R2 both included S22; however, despite being retrieved via database search, it was not included by the SLR-update. Similarly, for RQ3, if there were no false-negative results due to human judgment, our answer to this research question would be that differences between the conclusions from the SLR update and its replications as well as the conclusions using the superset of 15 studies did not differ to any larger extent.

*RQ5: What are the differences, if any, between the conclusions using the superset, and those in the original SLR?*

Concerning the original SLR, we have ten papers (see Table 1), including one study each: four studies show that WC has superior effort estimation accuracy (40%); four studies for CC and WC presenting similar effort estimation accuracy (40%), and two with inconclusive results (20%). The 15 papers in the superset, for the SLR update and its two replications, have in total 23 results. These 23 results are arranged as follows: nine results show that WC has superior effort estimation accuracy (39%); nine results for CC and WC presenting similar effort estimation accuracy (39%), two results show that CC has superior effort estimation accuracy (9%), and finally three results show inconclusive results (13%). The only relevant change observed is the two results showing that CC has superior effort estimation accuracy in comparison to WC; other than that, the percentages for the other three types of results are very similar.

## 5.2  Recommended guidelines based on a comparison of results

Table 5 shows very clearly that the only approach where all 15 studies were retrieved used: i) forward snowballing with a single iteration; ii) employs Google Scholar to find citations to a seed set containing all the primary studies included in the original SLR; and iii) also used the two papers describing the original SLR. As indicated in Section 3.1, under RQ2, it is preferable if the seed set helps to identify all papers in the superset. However, as shown in Table 6, it is also important to highlight that the volume of citations returned from Google Scholar (used by SLR-update-R1), and the volume of references returned from several database searches employed by the SLR-update, may increase the likelihood of false negatives, although many of the citations/references are easily discarded as irrelevant.

Furthermore, the volume of citations returned from Google Scholar is primarily related to the high number of citations to the two papers describing the original SLR, and we would argue that citations to the papers describing the original SLR could not be ignored. Moreover, there is a significant overlap in references, and hence, the number of unique references is substantially lower.

Concerning group-work, on the one hand, SLR-update-R1 was carried out by a single person. On the other hand, although the SLR-update had a team of people, the initial filtering of titles and abstracts was done separately and individually, and only the titles and abstracts chosen independently were combined and discussed during a joint meeting attended by most of the authors. Therefore, we argue that, whenever there are the resources available, the entire selection of studies should be made independently by, at least, two people, and then compared. We also argue that the best choice would be for these two people to be experienced in carrying out SLRs in SE. Although these two recommendations concerning the group-work configuration when carrying out SLRs in SE are not new and have been previously documented elsewhere (e.g. [17], [18] and [22]), we are reiterating the importance herein. In summary, we propose the following guidelines for updating SLR:

1. Use a seed set containing the original SLR + primary studies.

2. Use Google Scholar to search for papers.

3. Apply forward snowballing, without iteration, which ought to be sufficient since any paper published on the topic of an SLR should refer to either the SLR or at least one of the primary studies.

4. Include more than one researcher in the initial screening to minimize the risk of removing studies that should be included (false negatives).

It is important to note that our guidelines are based upon the approach that was found to be the most suitable. More specifically, the guidelines are based on the evidence gathered using the combination of the original SLR update and its two replications as input. The three updates were conducted with different mixtures of researchers and employing different search methods to identify primary studies. An evaluation of a separate SLR and its update is presented next to assess our guidelines. As other SLR updates are carried out in SE, by different groups of authors, and using different search strategies, further evidence can be gathered and used to support (or not) our recommended guidelines.

The research presented herein is focused on SLRs; however, we see no immediate reason as to why a similar approach could not also be suitable for mapping studies.

## 6. Evaluating the guidelines
Based on the design described in Section 3.2, the results from the remaining four steps in the evaluation process are presented in one subsection each here.

### 6.1 Preparation

#### 6.1.1 Supporting Excel sheets
Excel sheets were developed to support the different steps in the process. It includes worksheets for the alignment described below. Furthermore, given the decision to form different teams of two individuals, as described below, Excel sheets were developed to support the various teams. Thus, the Excel sheets were developed to support the distributed work.

#### 6.1.2 Synchronization of procedures
The first step in the search was to ensure that we used the same procedure in Google Scholar, i.e. find the same papers. This was particularly important since the authors were distributed and hence unable to meet physically. Thus, we first conducted a search to identify the number of papers citing the original SLR, i.e. forward snowballing. The search was conducted in Google Scholar using as input the title of the original SLR. Quotes and patents were unticked, and the search interval was set to 2012–2014. After some discussions about the use of Google Scholar, we agreed on the procedure and that the number of papers citing the original SLR, in the given time interval, was 68 papers. An Excel sheet with the 68 papers was created and included the links to the publications found through Google Scholar. Once this step was finished, we were ready to apply the inclusion criteria to each of the 68 papers.

#### 6.1.3 Alignment of inclusion
The next step was to ensure alignment in judgment. This was achieved by all four authors assessing and making recommendations concerning the inclusion or exclusion of each of the 68 papers citing the original SLR. There was a complete agreement for 51 papers (75%). The remaining 17 papers, for which there were disagreements, were further investigated. After discussions and looking at the majority (3-1 for inclusion or exclusion), decisions were made for 13 of the 17 papers (76%). Concerning the remaining four papers, the situation was as follows:

- In two cases, it was a draw, but after discussions and re-checking the inclusion/exclusion criteria, the four authors agreed.
- Two of the 68 papers identified were in reality, the same paper, despite using two different titles. The title was changed in a revision and hence published with a new title. Google Scholar found both versions. Thus, only the final version is included.
- One of the papers was an extended abstract and should therefore not be included, given that there was an exclusion criterion for such type of paper. This was missed by three of the four authors.

The assessment of the 68 papers served its purpose. It made us more aligned and aware of the mistakes made. In the end, 38 papers were included that cited the original SLR. Based on the assessment of the 68 papers citing the original SLR, it was agreed that we were sufficiently aligned to continue the assessment and divide the work between the researchers. This step clearly showed that we do make mistakes. Hence, it supports item 4 in the recommendations in Section 5.2, i.e. more than one researcher ought to conduct the screening of papers concerning inclusion and exclusion of papers. Therefore, it was agreed that we were ready to make the remaining assessment in teams of two (screening through citations to each of the remaining 90 papers in the seed set), rather than having all four researchers reviewing all citations to the 90 papers.

### 6.1.4 Team setup

It was decided to form teams of two individuals investigating the citations to the 90 primary studies. Instead of dividing into two teams and have 45 papers for each team, it was decided to mix the teams to avoid team and individual bias. Six teams were created so that each individual was in the same team as one of the other authors for 15 papers. For example, Author 1 formed a team with Author 2, Author 3 and Author 4, respectively for 15 papers each.

Furthermore, it was decided to only list a paper when it was found the first time, i.e. use the change of colours on the links when searching for citations in Google Scholar. Thus, each paper citing a paper in the seed set only has to be investigated once. The downside is that it was not possible to calculate the degree of agreement. However, it was judged to be essential to avoid bias, and to lessen the workload, in particular since the alignment work in the previous step had been successfully conducted. In summary, having six teams and working with different colleagues were judged as more important than being able to calculate the level of agreement.

### 6.2 Collecting data – Forward snowballing

Each team of two researchers assessed 15 of the 90 primary studies. There was a high level of agreement, and in cases of disagreements, the different assessments were quickly resolved by revisiting the inclusion criteria and discussing the papers. The differences were, in most cases related to mistakes, made by one of the reviewers. In some instances, it was a matter of interpreting whether the publication venue fulfilled the inclusion criteria, i.e. mostly a matter of checking whether the paper was peer-reviewed. In no instance, we needed to involve a third researcher to decide whether or not to include or exclude a paper. It should be noted that depending on the search date for citations to the primary studies, the number of papers found could vary somewhat.

Google Scholar provided slightly different results, although the end of the search interval was five years ago. When this happened, we looked at all papers found, i.e. even if a paper was only found by one of the researchers in the team. Given that the search was focused on papers published between 2012 and 2014, we did not expect this result using Google Scholar. However, it may be used to our advantage when using Google Scholar. Based on our experience, we recommend that searches ought to be done on different dates, preferably with days or even weeks apart. In this way, it is likely that more papers are found than if making the searches on the same day. Having six teams with different reviewers meant that in several cases the teams found the same papers, which means that duplicates had to be removed.

The original SLR includes some primary studies published in 2012, and hence papers published in 2012 and included in the original SLR needed to be removed since we did not want overlap between the original and the updated SLRs. Such overlap would not allow our result to represent a true update. In the end, 100 new papers (not included in the original SLR) were found that cited at least one of the primary studies and fulfilled the inclusion criteria. In total, it means that we identified 138 papers (38 citing the original SLR and 100 citing its primary studies) fulfilling the inclusion criteria, which were not included in the original SLR.

## 6.3 Results – Analysis of the SLR updates

### 6.3.1 Ensuring comparability between the SLR updates

There was a need to remove papers from 2012 included in the original SLR to make the SLR update by Manikas [32] comparable to our update. According to the updated SLR, it included 141 new papers, i.e. not included in the original SLR. When looking more closely at the list of 141 papers, it was observed that it included the original SLR [33]. Given that an updated SLR ought to list only new papers (primary studies) published after an SLR, we removed the original SLR from the number of papers found by Manikas. Thus, the total number of new papers included in the updated SLR was 140.

### 6.3.2 Comparisons between updates

Three research questions were formulated and presented in Section 3.2. The responses to the research questions provide important information concerning the proposed guidelines. It should be noted that the selected SLR was different from the SLR used to formulate the guidelines. Hence, the comparison and the conclusions based on the evaluation are important to assess the usefulness of the guidelines. The first research question (RQ6) for the evaluation is concerned with the papers in general. For the other two research questions (RQ7 and RQ8), a more in-depth analysis was needed. The objectives of the comparison are twofold. First, to investigate if the papers found in our analysis ought to have been found by Manikas (RQ7), and secondly, if there are any differences concerning the research contributions in the included papers (RQ8).

Thus, the two first authors evaluated the papers not included in both SLR updates independently, i.e. 33 papers found only by Manikas and 39 papers only found in our analysis. The evaluations were done independently and were later compared and discussed between them. The evaluation included investigating the source of the papers and listing the area of the research contribution.

The results of the research questions are as follows:

***RQ6:** How do the proposed guidelines perform in terms of identifying papers to be included in the updated SLR, in comparison to the published SLR update?*

The number of papers in common between the two updated SLRs (Manikas and ours) was 99 papers. Thus, on the one hand, Manikas has 41 papers found that were not found when applying the recommendations, in particular using forward snowballing; however, on the other hand, we had 39 papers that were not found by Manikas. The outcome requires further investigation. To ensure that we make a fair comparison, we decided to check whether the papers included by Manikas follow the inclusion criteria as strictly as interpreted in our analysis. If not, the two updates are not fully comparable. When going through the 41 papers included by Manikas, but not in our analysis, it was noted that eight of these did not meet the inclusion criteria. Four papers were book chapters; one paper was a summary of a keynote and three papers did not meet the inclusion criteria, although being published in the workshop series or one of the special issues. Our assumption was that Manikas looked at the workshop series and the two special issues, although ensuring that the papers still fulfilled the inclusion criteria with the requested wording being found in the title, abstract or keywords. However, this does not seem to be the case. Thus, these eight papers were excluded from further analysis. Therefore, the number of papers found by Manikas, and comparable to ours, was 132 papers. The papers included and the eight excluded papers in the analysis can be found in [61]. The overlap in percentage in the two cases was respectively 99/138 = 72% and 99/132 = 75%. Thus, the overlap is good, but further analysis of the non-overlapping papers is needed to understand the differences.

*RQ7: To what extent the papers found in our SLR update have also been found when conducting the first SLR update?*

We needed to investigate if the 39 papers found in our analysis, and not included by Manikas, ought to have been found by Manikas. For example, if the papers are available in the databases searched by Manikas, or in the special issues or workshop proceedings that were listed as studied by Manikas.

When analysing the 39 papers included by us, but not by Manikas, we identified two papers published in the special issue of the Journal of Systems and Software (and at least available in ScienceDirect) and another eight papers that are available in at least one of the five searched databases. The other 29 papers were not available in the databases searched by Manikas.

When looking at the inclusion criteria, it was expected that Manikas should have included these ten papers. However, the latter requires some further investigation, since it may be the case that the papers identified by us were not available in the databases when the study by Manikas was conducted. Thus, the ten papers not included by Manikas were analysed in some more detail concerning their availability. The outcome was as follows:

- Two papers related to the special issue in the Journal of Systems and Software on Software Ecosystems. This special issue was explicitly mentioned by Manikas, and hence these two papers should have been available, in particular since the special issue was published in 2012.
- Four papers were available through ACM Digital Library, although no online publication date is provided. However, three of the four papers were published at conferences in 2012 and at least they ought to have been available when Manikas conducted his SLR update. The fourth paper was published in 2014, and it may not have been available at the date of the analysis by Manikas.
- Two papers were published by Springer and listed as being online in 2013.
- Two papers were made available online after the publication of the updated SLR by Manikas. The papers were published online in October 2015, respectively April 2016.

Based on the analysis, it is concluded that seven of the ten papers ought to have been available in the databases when Manikas conducted the analysis. Nevertheless, even if not considering the remaining three papers, at least 29 papers identified in our update could not have been identified by Manikas, as his search was limited to specific databases.

*RQ8: Which of the two SLR updates identifies most papers focusing on research on the topic of the selected SLR?*

It was noted that some of the papers included by us fulfilled the inclusion criteria, but they did not present research concerning software ecosystems despite the term "software ecosystems" being used. For some papers, it was apparent that the term is used in general, for example, referring to a software ecosystem concerning a specific tool or open-source project. If assuming that the primary goal of an SLR ought to be to present research on a specific topic, it was decided that it was important to evaluate which papers contributed to research concerning software ecosystems. The analysis was only done for the papers found in either the update by Manikas or in our replication of the update, although using our recommendations concerning search strategy. It is done since the main objective is to investigate the search strategy and the papers in common between the two updates do not contribute to this objective. This means investigating whether or not each paper contributes to research concerning software ecosystems.

The outcome was as follows:

- For the 33 papers included by Manikas:
  o In no case, both evaluators regarded the paper as contributing to research concerning software ecosystems.
  o For three papers, at least one of the evaluators viewed the paper as contributing to research concerning software ecosystems. In this case, one evaluator listed two papers and one evaluator listed only one paper.
  o For the remaining 30 papers, none of the evaluators judged that the papers contributed to research concerning software ecosystems.
- For the 39 papers included by us:
  o For 21 papers, the evaluators agreed that the papers contributed to research concerning software ecosystems.
  o For eight papers, at least one of the evaluators viewed the paper as contributing to research concerning software ecosystems. In this case, one evaluator listed seven papers and one evaluator listed only one paper.
  o For the remaining ten papers, none of the evaluators judged that the papers contributed to research concerning software ecosystems.

Instead of discussing the disagreements or involving more evaluators, it was decided to make a conservative comparison, accepting a paper included by Manikas as contributing to research in the area of software ecosystems if at least one of the evaluators viewed it as a contributor to the area. For papers included by us, and not by Manikas, we decided to only count papers, after the independent evaluation of the paper, when both evaluators agreed that the paper contributed to research concerning software ecosystems. This choice was made to ensure that we did not favour our own update, and hence our recommended guidelines. Manikas has three papers out of 33 papers with research contributions concerning software ecosystems, i.e. this is equivalent to 3/33 = 9%. Concerning papers only included by us, the percentage becomes 21/39 = 54%.

Thus, we conclude that our update has better coverage of the research concerning software ecosystems.

In summary, the recommended guidelines provided better coverage of software ecosystems research than the procedure applied by Manikas, i.e. repeating the database search strategy in the original SLR.

## 6.4 Reporting – Observations from the evaluation

Based on the analysis, the following observations were made concerning the four items in the guidelines:

- **Item 1, seed set**: Using the original SLR and the primary studies as a seed set generated approximately as many papers as included in the SLR update by Manikas. At least in this particular case, the use of citations seems to provide more papers contributing to research concerning software ecosystems than focusing on searching for specific wordings in a database search.
- **Item 2, Google Scholar**: If having descriptive keywords to search for in an SLR, a database search ought to be very powerful. However, it needs to be complemented with a qualitative evaluation of the research contribution of the papers found. As shown here, the presence of certain words does not guarantee that the research contributes to a specific area. It is clear from the analysis on the research contribution above, where the recommendations outperformed the procedure applied in the SLR update by Manikas.
- **Item 3, forward snowballing**: Forward snowballing is based on citations, and papers contributing to a research area ought to refer to either an existing SLR or at least one of the primary studies included in an SLR. It was also clear, from the evaluation, that many research papers do not necessarily appear in the traditional databases (for example IEEEXplore, ACM, ScienceDirect and SpringerLink as used in the SLR update by Manikas) provided by different publishers. Thus, for SLR updates forward snowballing is an excellent alternative in comparison to redoing the search in databases for a different time interval. Snowballing may also be an excellent alternative for original SLRs given that many different publishers exist, and their papers may not appear in the databases mostly used when conducting SLRs. Thus, if looking for the best research sample on a research topic, then searches in databases ought to be at least complemented with, if not replaced by, searches using snowballing (both backward and forward snowballing). Thus, snowballing is a strong contender as a search strategy also for conducting an original SLR, whenever one has a good seed set. The key is to identify a seed set that is as good as possible. This can be achieved by using a hybrid approach [46], i.e. determining the seed set by searching a suitable database or an indexing service and then conducting snowballing from this dataset (as initially suggested in [45]).

- **Item 4, number of researchers**: We noticed some human mistakes when including and excluding papers. Thus, having more than one researcher evaluating papers for inclusion or exclusion was very valuable. Manikas is the sole author of the SLR update [32], and it may explain why some of the seven papers that we expected the SLR update should have included were missed. We also found the team structure of mixing teams with different pairs of reviewers very valuable since it made us more aligned, and we minimized both team and individual bias.

Given the good overlap between the two updates (in the range of 70–75%), the conclusions ought to be similar, although papers not fully focused on research concerning software ecosystems may skew the findings somewhat.

## 7. Threats to validity

**Formulation**: As previously stated, our recommended guidelines concerning how to best search for evidence when updating SLRs in SE are based on evidence gathered from investigating an original SLR + one update + two replications of the SLR update, which can be seen as a threat to conclusion validity. Our goal was to formulate our guidelines by using evidence from an existing SLR, its update and corresponding replications, rather than to carry out a formal experiment with a simulated scenario, or to re-do a few SLR updates ourselves. The chosen combination was the only one that had a range of different authors for the original SLR, its update and the replications, and where different search strategies were employed to identify primary studies. A separate evaluation has also been conducted and reported here to mitigate the threat to conclusion validity further.

The diversity of authors helped reduce bias when applying the different search methods to identify primary studies, and the use of different search strategies provided an opportunity for these to be compared; such comparison informed our guidelines. One of the authors in this paper (Mendes) has taken part in and knows the original SLR very well and has cited the original SLR in all the subsequent studies on the topic. Furthermore, many recent studies on the SLR topic were conducted with the participation of this paper's authors (not including Wohlin), who are well aware of (and cited) the original SLR. Therefore, it would seem that such knowledge could have influenced the effectiveness of the snowballing search in updating SLRs. However, studies from the other authors, retrieved using database searches, were also successfully retrieved using forward snowballing, which contradicts the "higher effectiveness" argument.

A potential threat towards the use of forward snowballing for updating systematic literature reviews is if a substantial number of new papers do neither cite the SLR to be updated nor the primary studies in the SLR. The situation may, for example, occur if an SLR is getting "very" old, and the area investigated has changed substantially over time. However, some foundational papers in the area ought to be cited, and hence the risk is judged to be low. Although, if this is suspected to be the case, we would suggest identifying a validation set of papers as a mitigation action. The validation set should include a collection of papers that are known by the authors and ought to be included in the updated SLR. If several of the papers in the validation set are citing neither the SLR nor the primary studies in the SLR, we would suggest that the SLR ought to be managed as a new SLR. When to update an SLR is further discussed in [42].

**Evaluation**: In this case, the threats to conclusion validity are primarily related to selection bias and evaluator bias. There is a risk that our selection of an SLR for the evaluation may bias the findings towards our guidelines. However, the inclusion criteria for the selected SLR are primarily focused on specific words. Hence, it ought to favour a database search or an indexing service search (such as Google Scholar) and not a citation search. Most inclusion criteria for SLRs regularly include some judgment based on the content, which is not the case for the selected SLR.

A potential threat in literature reviews is publication bias, i.e. papers with certain characteristics are more often published or more often retrieved. For example, there may be a tendency that papers with positive results are more easily published than those with negative results. However, it is probably least sensitive for forward snowballing, since it is a matter of what the papers, included in our update, cite, not that the papers are cited. Thus, forward snowballing depends on the visibility of the original SLR and the primary studies in the original SLR. The publication bias is higher for backward snowballing and database searches. For example, database searches are often done on databases serving some of the most well-established publishers, and hence the publication bias exists also when doing database searches. We found a number of papers not visible in the major databases, which points to that forward snowballing throws a broader net than searching in specific databases. In summary, any publication bias favours the guidelines over other approaches.

Furthermore, there is a risk that the individual researchers become biased, given that there is a vested interest in the guidelines. However, the selection of the SLR for the evaluation based on predefined criteria minimized this threat, and the composition of different teams mitigated the risk of both team and individual bias. Overall, it is judged that the design of the study and the predefined criteria for selecting an SLR to use in the evaluation help minimising the conclusion validity threats.

## 8. Conclusions

This paper investigates, proposes guidelines and evaluates an essential aspect concerning the update of SLRs in SE, i.e. "how" to best search for evidence to update an SLR. This aspect is addressed by using the results from a comparison of different search strategies (e.g., database search and forward snowballing processes) used by an SLR update and two replications, in the topic of effort estimation. The recommended guidelines are as follows:

1. Use a seed set containing the original SLR + primary studies.

2. Use Google Scholar to search for papers.

3. Apply forward snowballing, without iteration, which ought to be sufficient since any paper published on the topic of an SLR should refer to either the SLR or at least one of the primary studies.

4. Include more than one researcher in the initial screening to minimize the risk of removing studies that should be included (false negatives).

The guidelines have been evaluated using an SLR and its update. The evaluation supports the recommended guidelines. It is notable that the two investigated updates, i.e. for the formulation of the guidelines and for the evaluation of the guidelines, are different. For the formulation of the guidelines, an SLR with a narrow scope, and hence relatively few primary studies, was used. While for the evaluation, an SLR covering a broad area, and numerous primary studies, was used. Despite the differences, the suggested guidelines came out as superior in the evaluation. It is encouraging.

The guidelines are based on the logic that new papers ought to cite either the original SLR or, at least, one of the primary studies in the original SLR. Furthermore, the use of the guidelines on updating an SLR outperforms the original update when it comes to finding papers focusing on research on software ecosystems. Independently, we suggest that further investigations be carried out on how to update

SLRs in SE, for example, based on evidence from SLRs + updates, or even by carrying out formal experiments. Further studies and their findings will broaden our understanding in this area. They may provide support for how to effectively and efficiently conduct SLR updates, which will become an even more critical area since SLRs may need to be updated after some time to capture new development and research in a field. Thus, the findings and the guidelines should be seen as a promising way forward for updating SLRs.

## Acknowledgement

Dr. Katia Romero Felizardo is supported by a research grant from the Brazilian funding agency CNPq with reference number: 401033/2016-3. Furthermore, we would like to thank the anonymous reviewers and the guest editors for providing valuable feedback that helped improve the paper.